\documentclass[a4paper,11pt]{article}
\usepackage{pos}
\usepackage[english]{babel}
\usepackage{bm}
\usepackage{bbm}
\usepackage{mathtools}
\usepackage{graphicx}
\usepackage{tikz}
\usetikzlibrary{arrows,shapes,backgrounds,calc,positioning,tikzmark,patterns,automata,decorations.markings}
\tikzset{fontscale/.style={font=\relsize{#1}}}
\tikzset{->-/.style={decoration={
  markings,
  mark=at position #1 with {\arrow{>}}},postaction={decorate}}}
\tikzset{-<-/.style={decoration={
  markings,
  mark=at position #1 with {\arrow{<}}},postaction={decorate}}}

\tikzset{cross/.style={cross out,draw,minimum size=2*(#1-\pgflinewidth),inner sep=0pt, outer sep=0pt}}

\tikzset{
  pics/carc/.style args={#1:#2:#3}{
    code={
      \draw[pic actions] (0,0) -- (#1:#3) arc(#1:#2:#3) -- cycle;
    }
  }
}
\usepackage{picture}
\usepackage{cancel}
\usepackage{tabularx}
\usepackage{tensor}
\usepackage{float}
\usepackage{fancyhdr}
\usepackage{sidecap}
\usepackage{xcolor}
\usepackage{scalerel}
\usepackage{relsize}
\usepackage[absolute,overlay]{textpos}
\usepackage[customcolors]{hf-tikz}
\usepackage{cals}
\usepackage{enumitem}
\setlist{nolistsep}
\usepackage{pifont}
\usepackage[export]{adjustbox} 
\usepackage[most]{tcolorbox}
\usepackage{varwidth}
\usepackage{ulem}
\tcbuselibrary{skins}

\newtcolorbox{hlbox}[2][red]{
colbacktitle=#1!10,
colback=white!95!#1,
coltitle=black,
fonttitle=\bfseries,
colframe=#1!50,
boxrule=0.5pt,
titlerule=0pt,
title={\strut#2},
arc=3pt,
middle=0pt,
boxsep=0pt,
left skip=0pt,
right skip=0pt}

\let\originalleft\left
\let\originalright\right
\renewcommand{\left}{\mathopen{}\mathclose\bgroup\originalleft}
\renewcommand{\right}{\aftergroup\egroup\originalright}
\newcommand{\e}{\operatorname{e}}
\newcommand{\SU}[1]{\operatorname{SU}\left(#1\right)}

\newcommand{\Un}[1]{\operatorname{U}\left(#1\right)}

\newcommand{\su}[1]{\mathfrak{su}\left(#1\right)}

\newcommand{\of}[1]{\left(#1\right)}

\newcommand{\sof}[1]{\bigl(\big.#1\big.\bigr)}

\newcommand{\cof}[1]{\left\{\right.#1\left.\right\}}

\newcommand{\loint}[1]{\left(#1\right]}

\newcommand{\diag}{\operatorname{diag}}
\newcommand{\vspan}{\operatorname{span}}
\newcommand{\trace}{\operatorname{tr}}

\newcommand{\repart}{\operatorname{Re}}

\newcommand{\ii}{\mathrm{i}}
\newcommand{\idd}[2]{\mathrm{d}^{#2}#1}
\newcommand{\dd}{\mathrm{d}}

\newcommand{\order}[1]{\mathcal{O}\big(#1\big)}

\newcommand{\id}{\mathbbm{1}}

\renewcommand*\[{\begin{equation}}
\renewcommand*\]{\end{equation}}

\renewcommand*\hat[1]{\widehat{#1}}
\let\oldstackrel\stackrel
\renewcommand*\stackrel[2]{{\scriptstyle\oldstackrel{#1}{#2}}}

\definecolor{emphcol}{rgb}{1.,0,0}
\let\oldemph\emph
\renewcommand*\emph[1]{\oldemph{\textcolor{emphcol}{#1}}}

\newcommand{\psmatrix}[1]{\left(\begin{smallmatrix}#1\end{smallmatrix}\right)}
\newcommand{\tdiagm}[1]{\psmatrix{\e^{\ii\theta^{\of{1}}_{#1}} &  & \\
& \ddots & \\
& & \e^{\ii\theta^{\of{N}}_{#1}}}}
\newcommand{\tdiagmc}[1]{\psmatrix{\e^{-\ii\theta^{\of{1}}_{#1}} &  & \\
 & \ddots & \\
 & & \e^{-\ii\theta^{\of{N}}_{#1}}}}

\makeatletter
\pgfarrowsdeclare{open cap}{open cap}
{\pgfarrowsleftextend{+0pt}\pgfarrowsrightextend{+0.5\pgflinewidth}}
{
  \pgfmathsetlength{\pgfutil@tempdimb}{.5*\pgflinewidth-.5*\pgfinnerlinewidth}%
  \pgfsetlinewidth{\pgfutil@tempdimb}
  \pgfsetbuttcap
  \pgfsetdash{}{0pt}
  \pgfmathsetlength{\pgfutil@tempdima}{.5*\pgfutil@tempdimb+.5*\pgfinnerlinewidth}%
  \pgfpathmoveto{\pgfqpoint{0pt}{\pgfutil@tempdima}}
  \pgfpatharc{270}{90}{-\pgfutil@tempdima}
  \pgfusepathqstroke
}
\makeatother

\title{Bulk-preventing actions for SU(N) gauge theories}

\author*[a]{Tobias Rindlisbacher}
\author[b,c]{Kari Rummukainen}
\author[b,c]{Ahmed Salami}

\affiliation[a]{AEC, Institute for Theoretical Physics, University of Bern, Sidlerstrasse 5, CH-3012 Bern, Switzerland}

\affiliation[b]{Department of Physics,
P.O. Box 64, FI-00014 University of Helsinki, Finland}

\affiliation[c]{Helsinki Institute of Physics,
P.O. Box 64, FI-00014 University of Helsinki, Finland}

\emailAdd{trindlis@itp.unibe.ch}
\emailAdd{kari.rummukainen@helsinki.fi}
\emailAdd{ahmed.salami@helsinki.fi}

\abstract{We introduce a one-parameter family of SU(N) gauge actions which, when used in combination with an HMC update algorithm, prevent the gauge system from entering an artificial bulk-"phase". We briefly discuss the mechanism behind the bulk-prevention and present test results for different SU(N) gauge groups.}

\FullConference{%
 The 38th International Symposium on Lattice Field Theory, LATTICE2021
  26th-30th July, 2021
  Zoom/Gather@Massachusetts Institute of Technology
}

\begin{document}
\maketitle

\section{Intorduction}
Following Wilson's prescription, the lattice discretization of a $\SU{N}$ gauge theory is obtained by promoting the Lie algebra valued continuum gauge field, 
\[
A_{\mu}\of{x'}=\sum_{I} A^{I}_{\mu}\of{x'}\,T^{I}\quad,\quad \vspan\cof{T^{I}}_{I=1,\ldots,N^2-1}=\su{N}\ ,\label{eq:contgaugefield}
\] 
to Lie group valued link variables,
\[
U_{\mu}\of{x}\,=\,\mathcal{P}\e^{\ii\,\int_{a x}^{a\,(x+\hat{\mu})}\dd{x'}\,A_{\mu}\of{x'}}\,\in\,\SU{N}\  ,\label{eq:latgaugefield}
\]
which can be interpreted as the gauge parallel transporters along links from a site $x$ to a neighbouring site $x+\hat{\mu}$. The leading $\mathcal{P}$ on the right-hand side of \eqref{eq:latgaugefield} indicates that path ordering should be applied when evaluating the exponential of the line integral. The relation between the coordinate $x'\in\mathbb{R}^4$ in \eqref{eq:contgaugefield} and the coordinate $x\in\mathbb{Z}^4$ in \eqref{eq:latgaugefield} is given by $x=a\,x'$, where $a$ is the lattice spacing, and $\hat{\mu}$ refers to the unit vector in $\mu$-direction. Parallel transporters over longer distances are then expressed as product of consecutive link variables and a lattice gauge action can be defined in terms of link variables by requiring that in the limit $\of{a\to 0}$ the lattice gauge action converges to the continuum gauge action,
\[
S_{G}=\frac{1}{4\,g_0^2}\int\idd{x'}{4}\trace\of{F_{\mu\nu}\of{x'}F_{\mu\nu}\of{x'}}\ .
\]
Wilson proposed the gauge action~\cite{Wilson:1974sk}
\[
S_{G,W}=\frac{\beta}{N}\sum\limits_{x}\sum\limits_{\mu<\nu}\repart\trace\sof{\id-U_{\mu\nu}\of{x}}\ ,\label{eq:wilsongaugeaction}
\]
which, as is well known, satisfies the above condition and is here
written in terms of the inverse bare gauge coupling $\beta=2\,N/g_0^2$ and the plaquette variables
\[
U_{\mu\nu}\of{x}=U_{\mu}\of{x}\,U_{\nu}\of{x+\hat{\mu}}\,U^{\dagger}_{\mu}\of{x+\hat{\nu}}\,U^{\dagger}_{\nu}\of{x}\ .\label{eq:plaquettevariable}
\]
The gauge action \eqref{eq:wilsongaugeaction} and improved versions of it~\cite{Luscher:1984xn} are presumably the most well known and most used ones in Monte Carlo studies of $\SU{N}$ lattice gauge theories. They are, however, not unique and might not be the best choice for the study of lattice gauge theories at strong coupling, as they allow the gauge system to enter a so-called "bulk-phase". The latter is not necessarily a proper phase, but simply a region in parameter space of the lattice theory where lattice artefacts dominate in ensemble averages. As a consequence the relation between lattice and continuum results becomes very complicated or can even be lost completely, if bulk and continuum phase are separated by a first order transition.

\section{Avoiding the lattice bulk phase}\label{sec:avoidinglatbulkphase}

In this section we propose a characterization of "bulk configuration" in $\SU{N}$ lattice gauge systems, that allows for the definition of a family of lattice gauge actions, which separate such bulk configurations from regular ones by an infinite potential barrier, while still yielding the same naive continuum limit as Wilson's gauge action.  

\subsection{Motivation in $\Un{1}$}\label{ssec:motivationinu1}
\begin{figure}[htbp]
\centering
\begin{minipage}[t]{0.49\linewidth}
\vspace{0pt}
\centering
\includegraphics[width=0.85\linewidth,keepaspectratio]{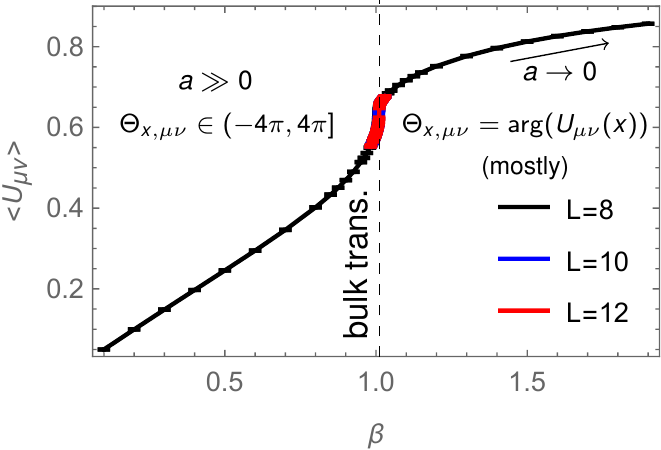}
\caption{The $\Un{1}$ lattice gauge theory with Wilson gauge action \eqref{eq:wilsongaugeactionu1} undergoes a bulk-transition at $\beta=\beta_b\approx 1$. For $\beta<\beta_b$, the system is in the bulk phase, where the lattice spacing, $a$, can be considered large and $\Theta_{x,\mu\nu}$ from \eqref{eq:totplaqphaseu1} explores the full $\loint{-4\pi,4\pi}$-interval. For $\beta>\beta_b$, the system is in the continuum phase, where the lattice spacing tends to zero if $\beta\to\infty$. In this phase, $\Theta_{x,\mu\nu}$ can still be outside the $\loint{-\pi,\pi}$-interval, but the fraction of such plaquettes quickly drops as $\beta$ is increased and one mostly has $\Theta_{x,\mu\nu}=\arg\of{U_{\mu\nu}\of{x}}$ . }
\label{fig:phasediagramu1}
\end{minipage}\hfill
\begin{minipage}[t]{0.49\linewidth}
\vspace{0pt}
\begin{enumerate}[label=(\alph*)]
\item\label{en:linkwrap} link wraps around $\loint{-\pi,\pi}$-interval:\\[-17pt]
\begin{center}
\begin{tikzpicture}[scale=0.24,nodes={inner sep=0},every node/.style={transform shape}]
  \def\tha{-155.0}
  \pgfmathparse{\tha<0 ? "blue" : "red"}
  \edef\cola{\pgfmathresult}
  \def\thb{-60.0}
  \pgfmathparse{\thb<0 ? "blue" : "red"}
  \edef\colb{\pgfmathresult}
  \def\thc{135.0}
  \pgfmathparse{\thc<0 ? "blue" : "red"}
  \edef\colc{\pgfmathresult}
  \def\thd{100.0}
  \pgfmathparse{\thd<0 ? "blue" : "red"}
  \edef\cold{\pgfmathresult}

  \pgfmathparse{\tha+\thb+\thc+\thd}
  \edef\thtot{\pgfmathresult}
  
  \pgfmathparse{\thtot<0 ? "blue" : "red"}
  \edef\coltot{\pgfmathresult}
  
  \def\thaalt{155.0}
  \pgfmathparse{\thaalt<0 ? "blue" : "red"}
  \edef\colaalt{\pgfmathresult}
  
  \pgfmathparse{\thaalt+\thb+\thc+\thd}
  \edef\thtotalt{\pgfmathresult}

  \pgfmathparse{\thtotalt<0 ? "blue" : "red"}
  \edef\coltotalt{\pgfmathresult}

  \pgfpointtransformed{\pgfpointxy{1}{1}};
  \pgfgetlastxy{\vx}{\vy};
  \begin{scope}[node distance=\vx and \vy,fontscale=-1]
    \draw[-<-=.18,thick,black] (0,0) -- (4,0);
    \draw[-<-=.18,thick,black] (4,0) -- (4,4);
    \draw[-<-=.18,thick,black] (4,4) -- (0,4);
    \draw[-<-=.18,thick,black] (0,4) -- (0,0);
    
    \draw node[black,below,scale=3.5] at (0,-0.2) {$x\phantom{\hat{\mu}}$};
    \draw node[black,below,scale=3.5] at (4.2,-0.2) {$x+\hat{\mu}$};
    \draw node[black,above,scale=3.5] at (5.1,4.2) {$x+\hat{\mu}+\hat{\nu}$};
    \draw node[black,above,scale=3.5] at (-0.2,4.2) {$x+\hat{\nu}\vphantom{\hat{\mu}}$};
    \draw[thick,fill=white] (2,0) circle (0.8) pic[\cola,fill=\cola!50!white]{carc=90:90-\tha:0.8};  
    \draw[thick,fill=white] (4,2) circle (0.8) pic[\colb,fill=\colb!50!white]{carc=90:90-\thb:0.8};
    \draw[thick,fill=white] (2,4) circle (0.8) pic[\colc,fill=\colc!50!white]{carc=90:90-\thc:0.8};    
    \draw[thick,fill=white] (0,2) circle (0.8) pic[\cold,fill=\cold!50!white]{carc=90:90-\thd:0.8};
    \draw[thick,fill=white] (2,1.9) circle (0.8) pic[\coltot,fill=\coltot!50!white,->]{carc=90:90-\thtot:0.8};
    \draw[black,fill=white] (0,0) circle (5pt); 
    \draw[black,fill=white] (4,0) circle (5pt);
    \draw[black,fill=white] (4,4) circle (5pt);
    \draw[black,fill=white] (0,4) circle (5pt);

    \def\xdisp{10}
    \draw node[black,scale=4.5] at (2+0.5*\xdisp,2.0) {$\Longrightarrow$};    
    
    \draw[-<-=.18,thick,black] (\xdisp,0) -- (4+\xdisp,0);
    \draw[-<-=.18,thick,black] (4+\xdisp,0) -- (4+\xdisp,4);
    \draw[-<-=.18,thick,black] (4+\xdisp,4) -- (\xdisp,4);
    \draw[-<-=.18,thick,black] (\xdisp,4) -- (\xdisp,0);
    
    \draw node[black,below,scale=3.5] at (\xdisp,-0.2) {$x\phantom{\hat{\mu}}$};
    \draw node[black,below,scale=3.5] at (4.2+\xdisp,-0.2) {$x+\hat{\mu}$};
    \draw node[black,above,scale=3.5] at (5.1+\xdisp,4.2) {$x+\hat{\mu}+\hat{\nu}$};
    \draw node[black,above,scale=3.5] at (\xdisp-0.2,4.2) {$x+\hat{\nu}\vphantom{\hat{\mu}}$};
    \draw[thick,fill=white] (2+\xdisp,0) circle (0.8) pic[\colaalt,fill=\colaalt!50!white]{carc=90:90-\thaalt:0.8};  
    \draw[thick,fill=white] (4+\xdisp,2) circle (0.8) pic[\colb,fill=\colb!50!white]{carc=90:90-\thb:0.8};
    \draw[thick,fill=white] (2+\xdisp,4) circle (0.8) pic[\colc,fill=\colc!50!white]{carc=90:90-\thc:0.8};    
    \draw[thick,fill=white] (\xdisp,2) circle (0.8) pic[\cold,fill=\cold!50!white]{carc=90:90-\thd:0.8};
    \draw[thick,fill=white] (2+\xdisp,1.9) circle (0.8) pic[\coltotalt,fill=\coltotalt!50!white,->]{carc=90:90-\thtotalt:0.8};
    \draw[black,fill=white] (\xdisp,0) circle (5pt); 
    \draw[black,fill=white] (4+\xdisp,0) circle (5pt);
    \draw[black,fill=white] (4+\xdisp,4) circle (5pt);
    \draw[black,fill=white] (\xdisp,4) circle (5pt); 
  \end{scope}
\end{tikzpicture}\\[-20pt]
\end{center}
\item\label{en:nolinkwrap} $\Theta_{x,\mu\nu}$ grows continuously:\\[-17pt]
\begin{center}
\begin{tikzpicture}[scale=0.24,nodes={inner sep=0},every node/.style={transform shape}]
  \def\tha{20.0}
  \pgfmathparse{\tha<0 ? "blue" : "red"}
  \edef\cola{\pgfmathresult}
  \def\thb{40.0}
  \pgfmathparse{\thb<0 ? "blue" : "red"}
  \edef\colb{\pgfmathresult}
  \def\thc{50.0}
  \pgfmathparse{\thc<0 ? "blue" : "red"}
  \edef\colc{\pgfmathresult}
  \def\thd{60.0}
  \pgfmathparse{\thd<0 ? "blue" : "red"}
  \edef\cold{\pgfmathresult}

  \pgfmathparse{\tha+\thb+\thc+\thd}
  \edef\thtot{\pgfmathresult}
  
  \pgfmathparse{\thtot<0 ? "blue" : "red"}
  \edef\coltot{\pgfmathresult}
  
  \def\thaalt{40.0}
  \pgfmathparse{\thaalt<0 ? "blue" : "red"}
  \edef\colaalt{\pgfmathresult}
  
  \pgfmathparse{\thaalt+\thb+\thc+\thd}
  \edef\thtotalt{\pgfmathresult}
  
  \pgfmathparse{\thtotalt<0 ? "blue" : "red"}
  \edef\coltotalt{\pgfmathresult}

  \pgfpointtransformed{\pgfpointxy{1}{1}};
  \pgfgetlastxy{\vx}{\vy};
  \begin{scope}[node distance=\vx and \vy,fontscale=-1]
    \draw[-<-=.18,thick,black] (0,0) -- (4,0);
    \draw[-<-=.18,thick,black] (4,0) -- (4,4);
    \draw[-<-=.18,thick,black] (4,4) -- (0,4);
    \draw[-<-=.18,thick,black] (0,4) -- (0,0);
    
    \draw node[black,below,scale=3.5] at (0,-0.2) {$x\phantom{\hat{\mu}}$};
    \draw node[black,below,scale=3.5] at (4.2,-0.2) {$x+\hat{\mu}$};
    \draw node[black,above,scale=3.5] at (5.1,4.2) {$x+\hat{\mu}+\hat{\nu}$};
    \draw node[black,above,scale=3.5] at (-0.2,4.2) {$x+\hat{\nu}\vphantom{\hat{\mu}}$};
    \draw[thick,fill=white] (2,0) circle (0.8) pic[\cola,fill=\cola!50!white]{carc=90:90-\tha:0.8};  
    \draw[thick,fill=white] (4,2) circle (0.8) pic[\colb,fill=\colb!50!white]{carc=90:90-\thb:0.8};
    \draw[thick,fill=white] (2,4) circle (0.8) pic[\colc,fill=\colc!50!white]{carc=90:90-\thc:0.8};    
    \draw[thick,fill=white] (0,2) circle (0.8) pic[\cold,fill=\cold!50!white]{carc=90:90-\thd:0.8};
    \draw[thick,fill=white] (2,1.9) circle (0.8) pic[\coltot,fill=\coltot!50!white,->]{carc=90:90-\thtot:0.8};
    \draw[black,fill=white] (0,0) circle (5pt); 
    \draw[black,fill=white] (4,0) circle (5pt);
    \draw[black,fill=white] (4,4) circle (5pt);
    \draw[black,fill=white] (0,4) circle (5pt);

    \def\xdisp{10}
    \draw node[black,scale=4.5] at (2+0.5*\xdisp,2.0) {$\Longrightarrow$};    
    
    \draw[-<-=.18,thick,black] (\xdisp,0) -- (4+\xdisp,0);
    \draw[-<-=.18,thick,black] (4+\xdisp,0) -- (4+\xdisp,4);
    \draw[-<-=.18,thick,black] (4+\xdisp,4) -- (\xdisp,4);
    \draw[-<-=.18,thick,black] (\xdisp,4) -- (\xdisp,0);
    
    \draw node[black,below,scale=3.5] at (\xdisp,-0.2) {$x\phantom{\hat{\mu}}$};
    \draw node[black,below,scale=3.5] at (4.2+\xdisp,-0.2) {$x+\hat{\mu}$};
    \draw node[black,above,scale=3.5] at (5.1+\xdisp,4.2) {$x+\hat{\mu}+\hat{\nu}$};
    \draw node[black,above,scale=3.5] at (\xdisp-0.2,4.2) {$x+\hat{\nu}\vphantom{\hat{\mu}}$};
    \draw[thick,fill=white] (2+\xdisp,0) circle (0.8) pic[\colaalt,fill=\colaalt!50!white]{carc=90:90-\thaalt:0.8};  
    \draw[thick,fill=white] (4+\xdisp,2) circle (0.8) pic[\colb,fill=\colb!50!white]{carc=90:90-\thb:0.8};
    \draw[thick,fill=white] (2+\xdisp,4) circle (0.8) pic[\colc,fill=\colc!50!white]{carc=90:90-\thc:0.8};    
    \draw[thick,fill=white] (\xdisp,2) circle (0.8) pic[\cold,fill=\cold!50!white]{carc=90:90-\thd:0.8};
    \draw[thick,fill=white] (2+\xdisp,1.9) circle (0.8) pic[\coltotalt,fill=\coltotalt!50!white,->]{carc=90:90-\thtotalt:0.8};
    \draw[black,fill=white] (\xdisp,0) circle (5pt); 
    \draw[black,fill=white] (4+\xdisp,0) circle (5pt);
    \draw[black,fill=white] (4+\xdisp,4) circle (5pt);
    \draw[black,fill=white] (\xdisp,4) circle (5pt); 
  \end{scope}
\end{tikzpicture}
\end{center}
\end{enumerate}
\caption{The figure illustrates the two different ways in which $\Theta_{x,\mu\nu}$ form \eqref{eq:totplaqphaseu1} can grow beyond the $\loint{-\pi,\pi}$-interval: a) a single link (between $x$ and $x+\hat{\mu}$) winds around $\loint{\pi,\pi}$ which adds almost $2\,\pi$ to $\Theta_{x,\mu\nu}$; b) no link wraps around the $\loint{-\pi,\pi}$-interval, $\Theta_{x,\mu\nu}$ grows continuously bigger than $\pi$.}
\label{fig:contvsdiscontwindingu1}
\end{minipage}
\end{figure}

In the $\Un{1}$ case, the Wilson gauge action in \eqref{eq:wilsongaugeaction} reduces to
\[
S_{G,W}=\beta\,\sum\limits_{x}\sum\limits_{\mu<\nu}\repart\sof{\id-U_{\mu\nu}\of{x}}\ ,\label{eq:wilsongaugeactionu1}
\]
and the Abelian link variables can be written as
\[
U_{\mu}\of{x}=\e^{\ii\,\theta_{x,\mu}}\quad\text{with}\quad \theta_{x,\mu}=a\,A_{\mu}\of{x}\,\in\,\loint{-\pi,\pi}\ .\label{eq:abelianlinkvar}
\]
Let us now define,
\[
\Theta_{x,\mu\nu}=\theta_{x,\mu}+\theta_{x+\hat{\mu},\nu}-\theta_{x+\hat{\nu},\mu}-\theta_{x,\nu}\,\in\,\loint{-4\pi,4\pi}\ ,\label{eq:totplaqphaseu1}
\]
and note that while $\Theta_{x,\mu\nu}$ in \eqref{eq:totplaqphaseu1} can vary in the interval $\loint{-4\pi,4\pi}$, the gauge action \eqref{eq:wilsongaugeactionu1} depends only on
\[
\arg\of{U_{\mu\nu}\of{x}}\,\in\,\loint{-\pi,\pi}\ .\label{eq:argplaqu1}
\]
As illustrated in Fig.~\ref{fig:phasediagramu1}, the gauge action \eqref{eq:wilsongaugeactionu1} produces a bulk-transition at $\beta=\beta_b\approx 1$. For $\beta<\beta_b$, the system is in the bulk phase, where the lattice spacing, $a$, can be considered large and $\Theta_{x,\mu\nu}$ from \eqref{eq:totplaqphaseu1} explores the full $\loint{-4\pi,4\pi}$-interval. For $\beta>\beta_b$, the system is in the continuum phase, where the lattice spacing tends to zero if $\beta\to\infty$. In this phase, $\Theta_{x,\mu\nu}$ can still be outside the $\loint{-\pi,\pi}$-interval, but the fraction of such plaquettes quickly drops as $\beta$ is increased and most of the time, one has that $\Theta_{x,\mu\nu}=\arg\of{U_{\mu\nu}\of{x}}$ . 

The fact that plaquettes with $\Theta_{x,\mu\nu}\notin\loint{-\pi,\pi}$ also appear in the continuum phase indicates, that the value of $\Theta_{x,\mu\nu}$ by itself cannot be used to distinguish bulk- from continuum-configurations. However, as illustrated in Fig.~\ref{fig:contvsdiscontwindingu1}, there are two qualitatively different ways in which plaquettes with $\Theta_{x,\mu\nu}\notin\loint{-\pi,\pi}$ can be produced when starting from a configuration in which initially $\Theta_{x,\mu\nu}=\arg\of{U_{\mu\nu}\of{x}}$ for all plaquettes: 
\begin{enumerate}[label=(\alph*)]
\item\label{en:caselinkwrap} one of the links of the plaquette can move across the boundary of the $\loint{-\pi,\pi}$-interval and wrap around, which adds roughly $\pm 2\,\pi$ to $\Theta_{x,\mu\nu}$. As indicated in part~\ref{en:linkwrap} of Fig.~\ref{fig:contvsdiscontwindingu1}, the latter can happen also if $\arg\of{U_{\mu\nu}\of{x}}$ is close to $0$. Also it should be noted that such a wrapping link will produce a shift of almost $\pm 2\,\pi$ in the $\Theta_{x,\mu\nu}$ of all the plaquettes that contain this link.
\item\label{en:casenolinkwrap} No link wraps around the $\loint{-\pi,\pi}$-interval, but $\Theta_{x,\mu\nu}$ is already close to the boundary of the $\loint{-\pi,\pi}$-interval and finally crosses it. As indicated in part~\ref{en:nolinkwrap} of Fig.~\ref{fig:contvsdiscontwindingu1}, this can only happen if the gauge action cannot sufficiently prevent $\arg\of{U_{\mu\nu}\of{x}}$ from growing close to $\pm\pi$. As this way of "plaquette wrapping" does not involve the wrapping of individual links, it can happen to single plaquettes, and can cause local minima in the gauge action.  
\end{enumerate}
As case~\ref{en:casenolinkwrap} can only occur if individual plaquette angles are allowed to grow close to $\pm\pi$, which corresponds to the maximum value of the local action of a plaquette, this case is likely to occur only in the bulk phase, where $\beta$ is too small and the gauge action cannot grow sufficiently large to oppose the entropy-driven randomization of the links and plaquettes in the lattice system. We will therefore introduce in Sec.~\ref{ssec:bulkprevaction} a family of actions, which will prevent plaquette wrappings of type~\ref{en:casenolinkwrap}. As it turns out, this is sufficient to get rid of the bulk-transition. First, however, we discuss in Sec.~\ref{ssec:situationinsun} how the plaquette wrapping types \ref{en:caselinkwrap} and \ref{en:casenolinkwrap} generalize to the case of non-Abelian $\SU{N}$ lattice gauge theories.

\subsection{Situation in $\SU{N}$}\label{ssec:situationinsun}
In order to generalize the discussion from the previous section to $\SU{N}$, we diagonalize the link variables:
\[
U_{\mu}\of{x}=V_{x,\mu}^{\dagger}\diag\sof{\e^{\ii\theta^{\of{1}}_{x,\mu}},\ldots,\e^{\ii\theta^{\of{N}}_{x,\mu}}}V_{x,\mu}\quad,\quad \text{where}\quad \sum_{n=1}^{N}\theta^{\of{n}}_{x,\mu}=0\ ,
\]
and do the same with the plaquette variables:
\[
\begin{tikzpicture}[scale=0.9,baseline=(U.base),nodes={inner sep=0},every node/.style={transform shape}]
  \pgfpointtransformed{\pgfpointxy{1}{1}};
  \pgfgetlastxy{\vx}{\vy}
  \begin{scope}[node distance=\vx and \vy]
    \def\msize{4}
    \draw[-<-=.1,thick,black] (0,0) -- (\msize,0) node[fill=white,pos=0.5,scale=0.85] {$\tdiagm{{x},\mu}$};
    \draw[-<-=.1,thick,black] (\msize,0) -- (\msize,0.9*\msize) node[fill=white,pos=0.5,scale=0.85] {$\tdiagm{x+\hat{\mu},\nu}$};
    \draw[-<-=.1,thick,black] (\msize,0.9*\msize) -- (0,0.9*\msize) node[fill=white,pos=0.5,scale=0.85] {$\tdiagmc{x+\hat{\nu},\mu}$};
    \draw[-<-=.1,thick,black] (0,0.9*\msize) -- (0,0) node[fill=white,pos=0.5,scale=0.85] {$\tdiagmc{x,\nu}$};
    
    \draw[black,fill=white] (0,0) circle (\msize/20); 
    \draw[black,fill=white] (\msize,0) circle (\msize/20);
    \draw[black,fill=white] (\msize,0.9*\msize) circle (\msize/20);
    \draw[black,fill=white] (0,0.9*\msize) circle (\msize/20); 
    
    \draw node[fill=white,anchor=north east,scale=1,align=center] at (0,0) {$V^{\vphantom{\dagger}}_{x,\vphantom{\hat{\mu}}\nu}V^{\dagger}_{x,\vphantom{\hat{\mu}}\mu}$};
    \draw node[fill=white,anchor=north west,scale=1,align=center] at (\msize,0) {$V^{\vphantom{\dagger}}_{x,\vphantom{\hat{\mu}}\mu}V^{\dagger}_{x+\hat{\mu},\nu}$};
    \draw node[fill=white,anchor=south west,scale=1,align=center] at (\msize,0.9*\msize) {$V^{\vphantom{\dagger}}_{x+\hat{\mu},\nu}V^{\dagger}_{x+\hat{\nu},\mu}$};
    \draw node[fill=white,anchor=south east,scale=1,align=center] at (0,0.9*\msize) {$V^{\vphantom{\dagger}}_{x+\hat{\nu},\mu}V^{\dagger}_{x,\vphantom{\hat{\mu}}\nu}$};
    \draw node[fill=white,scale=1.1,anchor=east] (U) at (-0.4*\msize,0.9*\msize/2) {$U_{\mu\nu}\of{x}=V^{\dagger}_{x,\mu\nu}\tdiagm{x,\mu\nu}V^{\vphantom{\dagger}}_{x,\mu\nu}=$};
  \end{scope}
\end{tikzpicture}\ \label{eq:plaqdiag}
\]
where again, $\sum_{n=1}^{N}\,\theta_{x,\mu\nu}^{\of{n}}=0$ . As the products of $V_{x,\mu}$-matrices, appearing in \eqref{eq:plaqdiag} after the last equality sign, mix the eigenvalues of the link variables, the phases of plaquette eigenvalues cannot be represented as a simple sum of individual link-eigenvalue phases. Nevertheless, it is still true that each of the plaquette eigenvalues can leave the $\loint{-\pi,\pi}$-interval either by obtaining a large shift due to the wrapping of a link, analogous to case~\ref{en:caselinkwrap} of $\Un{1}$-discussion in Sec.~\ref{ssec:motivationinu1}, or by approaching and crossing the $\pm\pi$-boundary "smoothly" as in case~\ref{en:casenolinkwrap}. The latter case can be avoided, by preventing the plaquette eigenvalues from approaching the value $-1$. 

\subsection{Bulk-preventing action}\label{ssec:bulkprevaction}
To prevent plaquettes from having eigenvalues close to $-1$, we introduce the following family of gauge actions ($n\geq 1$):
\[
S_{G,b}=\frac{2\,\gamma}{n\,N}\sum\limits_{x}\sum\limits_{\mu<\nu}\trace\sof{\sof{\Omega^{\dagger}_{\mu\nu}\of{x}\Omega_{\mu\nu}\of{x}}^{-n}-\id}\quad,\quad\text{with}\quad \Omega_{\mu\nu}\of{x}=\sof{\id+{\underbrace{U_{\mu\nu}\of{x}}_{\mathclap{\text{plaquette}}}}}/2 .\label{eq:bpaction}
\]
The form of the actions \eqref{eq:bpaction} was inspired by the dislocation-prevention action introduced in~\cite{DeGrand:2014rwa}. The naive continuum limit of \eqref{eq:bpaction} is the same as for the Wilson gauge action $S_G$. This can be seen by writing $U_{\mu\nu}\of{x}=\exp\sof{\ii\,{\color{blue}s}\,F'_{\mu\nu}\of{x}}$, with ${\color{blue}s}\,F'_{\mu\nu}={\color{red}a^2}\,F_{\mu\nu}+\order{{\color{red}a^3}}$ and expanding two action in power series in $s$. One then finds:
\[
\frac{2}{n}\trace\sof{\sof{\Omega^{\dagger}_{\mu\nu}\of{x}\Omega_{\mu\nu}\of{x}}^{-n}-\id}=\repart\trace\of{\id-U_{\mu\nu}\of{x}}+{\color{blue}s^4}\frac{1+n}{8}\sof{\trace\sof{F'_{\mu\nu}\of{x}F'_{\mu\nu}\of{x}}/2}^2+\order{{\color{blue}s^6}}\ ,\label{eq:expansionbpaction}
\]
which means that for $\of{a\to 0}$, which implies also $\of{s\to 0}$, the two actions, \eqref{eq:bpaction} and \eqref{eq:wilsongaugeaction} are equivalent.

The actions \eqref{eq:bpaction} introduce an infinite potential barrier between bulk and continuum configurations. This is sufficient to ensure, that if we start a simulation from a cold configuration (all link variables equal to the identity) and use a hybrid Monte carlo (HMC) algorithm to update the gauge system, no bulk configurations will be produces.

One could infer, that this procedure yields a non-ergodic update algorithm. But, one should keep in mind, that the part of the configuration space that is not sampled, is irrelevant for the continuum limit of the theory. The algorithm prevents ensemble averages of the lattice system from being contaminated (or even dominated) by bulk-configurations. This should allow one to extract continuum physics also at stronger coupling. The same effect could be achieved by defining a modified measure, which gives zero weight to bulk-configurations. This would, however, be rather difficult to implement as the latter are hard to identify, once they are created. The use of an action \eqref{eq:bpaction} in combination with an HMC algorithm is a proxy to achieve the same effect but in a simpler and more economic way.

For $\Un{1}$ and $\SU{2}$, the actions in \eqref{eq:bpaction} have a similar effect as the topological actions discussed in \cite{Akerlund:2015zha,Bornyakov:1991gq,Nogradi:2018ivi}: the larger the inverse bare coupling $\gamma$, the stronger the plaquette values are repelled from $-1$ resp. $-\id$. For $\SU{N}$ with $N>2$ the effect of \eqref{eq:bpaction} is different from the one of the topological action discussed in \cite{Nogradi:2018ivi}, as for $N>2$ the trace of the plaquette does no longer completely determine the plaquette eigenvalues. However, an action which can have a similar effect as our actions \eqref{eq:bpaction} has been given in \cite{Brandt:2019ukx}. While for us it was desirable that the actions \eqref{eq:bpaction} do not prevent topology from fluctuating, there have also been attempts to find actions which keep the topology fixed \cite{Bietenholz:2005rd}.  

\section{Results}\label{sec:rsults}
To see whether the bulk-preventing actions~\eqref{eq:bpaction} deserve their name and how well they are able to reproduce the same weak coupling results as the Wilson gauge action, we carried out simulations with pure gauge $\SU{2}$, pure gauge $\SU{5}$ and $\SU{3}$ with $N_{f}=4$ Wilson fermion flavours. With the Wilson gauge action, all three of these theories enter a bulk phase for sufficiently small values of the inverse gauge coupling $\beta$. For pure gauge $\SU{2}$, the transition from continuum- to bulk-phase is of second order, while for pure gauge $\SU{5}$ and the fermionic $\SU{3}$-theory with sufficiently large $\kappa$, the transition is of first order. In the following we will discuss the three cases separately. We use the version of \eqref{eq:bpaction} with $n=2$. 
As according to \eqref{eq:expansionbpaction}, the Wilson gauge action \eqref{eq:wilsongaugeaction} and bulk-preventing action \eqref{eq:bpaction} agree only to order $\order{s^2}\sim \order{a^4}$, the inverse bare couplings $\beta$ and $\gamma$ will not be equal in the weak coupling limit, but in general be related by a relative shift.

\subsection{$\SU{2}$ pure gauge}\label{ssec:puregaugesu2results}

\begin{figure}[htbp]
\begin{minipage}[t]{0.51\linewidth}
\centering
\begin{minipage}[t]{0.49\linewidth}
\vspace{0pt}
\centering
\includegraphics[height=1.0\linewidth,keepaspectratio,right]{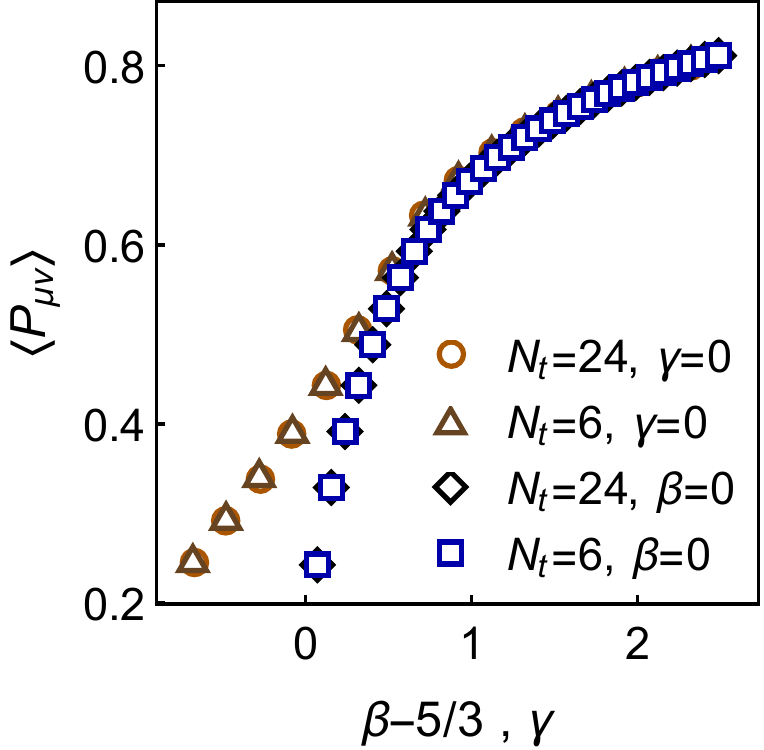}
\end{minipage}\hfill
\begin{minipage}[t]{0.49\linewidth}
\vspace{0pt}
\centering
\includegraphics[height=1.0\linewidth,keepaspectratio,right]{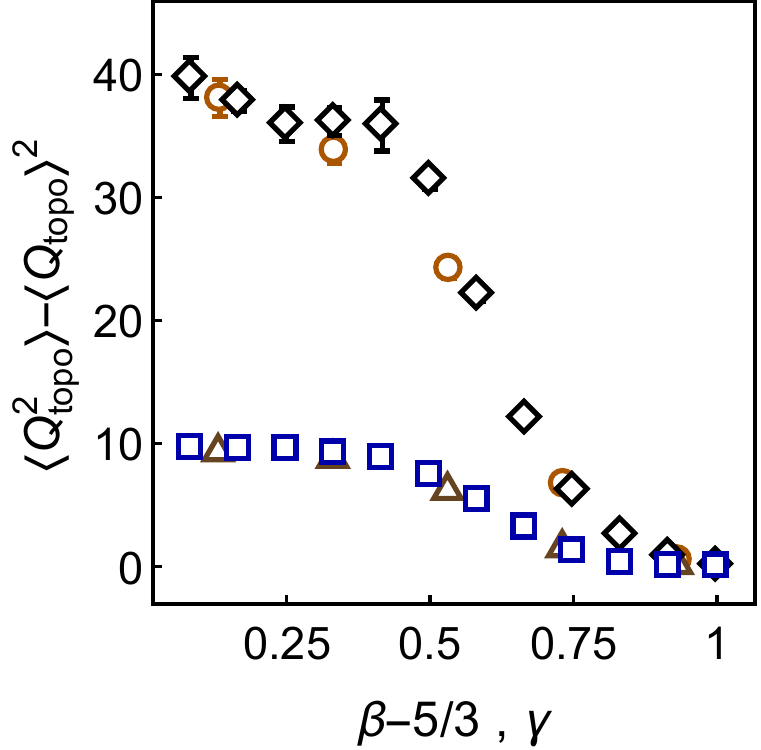}
\end{minipage}
\end{minipage}\hfill
\begin{minipage}[t]{0.51\linewidth}
\centering
\begin{minipage}[t]{0.49\linewidth}
\vspace{0pt}
\centering
\includegraphics[height=1.0\linewidth,keepaspectratio,right]{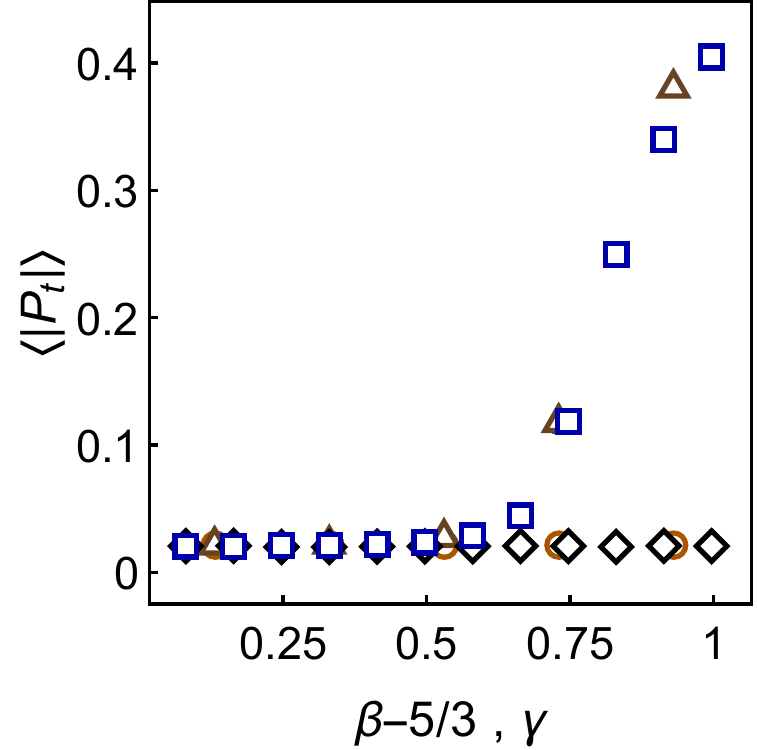}
\end{minipage}\hfill
\begin{minipage}[t]{0.49\linewidth}
\vspace{0pt}
\centering
\includegraphics[height=1.0\linewidth,keepaspectratio,right]{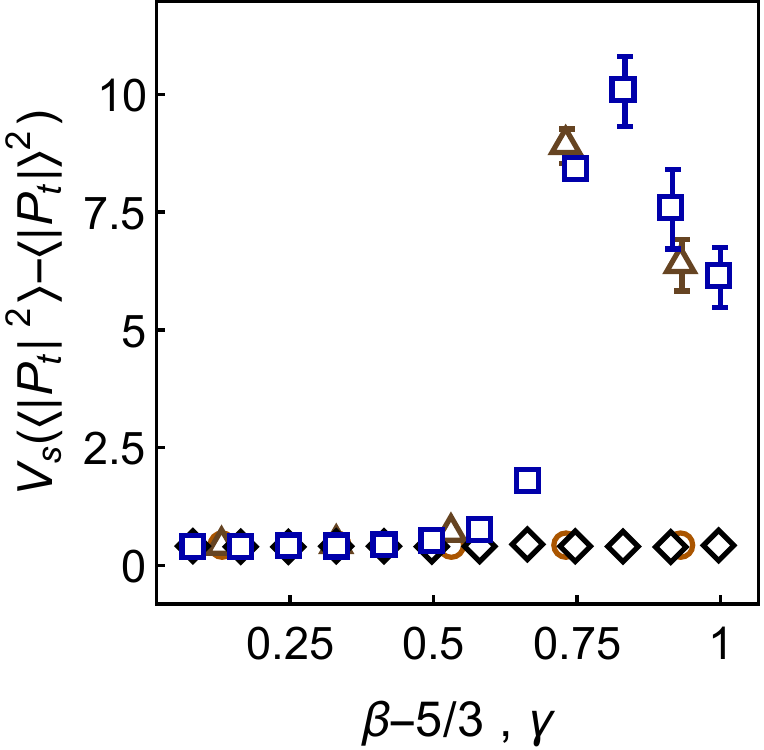}
\end{minipage}
\end{minipage}
\caption{The figure compares for pure gauge $\SU{2}$ results for average plaquette (1st column), topological susceptibility (2nd column), temporal Polyakov loop (3rd column) and temporal Polyakov loop variance (4th column), as obtained with the Wilson gauge action \eqref{eq:wilsongaugeaction} (orange circles and brown triangles) and the bulk-preventing action \eqref{eq:bpaction} with $n=2$ (black diamonds and blue squares). Plot legends are shown in the first column; $\gamma=0$ indicates that only the Wilson action was used to create the corresponding data, while $\beta=0$ indicates that only the bulk-preventing action was used. $N_t=6$ refers to finite temperature, while $N_t=24$ represents the zero-temperature case. The results are shown as function of $\beta-5/3$ resp. $\gamma$.}
\label{fig:actioncompsu2}
\end{figure}

In Fig.~\ref{fig:actioncompsu2} we compare results for pure gauge $\SU{2}$, obtained with the Wilson gauge (WG) action \eqref{eq:wilsongaugeaction} and the bulk-preventing (BP) action \eqref{eq:bpaction}. The data is plotted as function of $\beta-5/3$ (WG) and $\gamma$ (BP). The shift in $\beta$ has been determined by requiring that the spatial deconfinement transition occurs for the two actions at the same value of shifted $\beta$ and $\gamma$.

The first column in Fig.~\ref{fig:actioncompsu2} shows the average plaquette and we note that the shifted $\beta$ and $\gamma$ let the plaquette values for the two different actions agree at sufficiently weak coupling. Only below $\gamma=\beta-5/3\approx 1$, where the Wilson action enters the bulk-phase, the plaquette for the Wilson action starts  to deviate. The second column shows the topological susceptibility, which agrees very well for the two actions at weak and at strong coupling. This confirms that our BP action does not block processes which are relevant for topology changes. The last two columns show the temporal Polyakov loop and its variance, which also agree well for the two actions.

\subsection{$\SU{5}$ pure gauge}\label{ssec:puregaugesu5results}

\begin{figure}[htbp]
\begin{minipage}[t]{0.51\linewidth}
\centering
\begin{minipage}[t]{0.49\linewidth}
\vspace{0pt}
\centering
\includegraphics[height=1.0\linewidth,keepaspectratio,right]{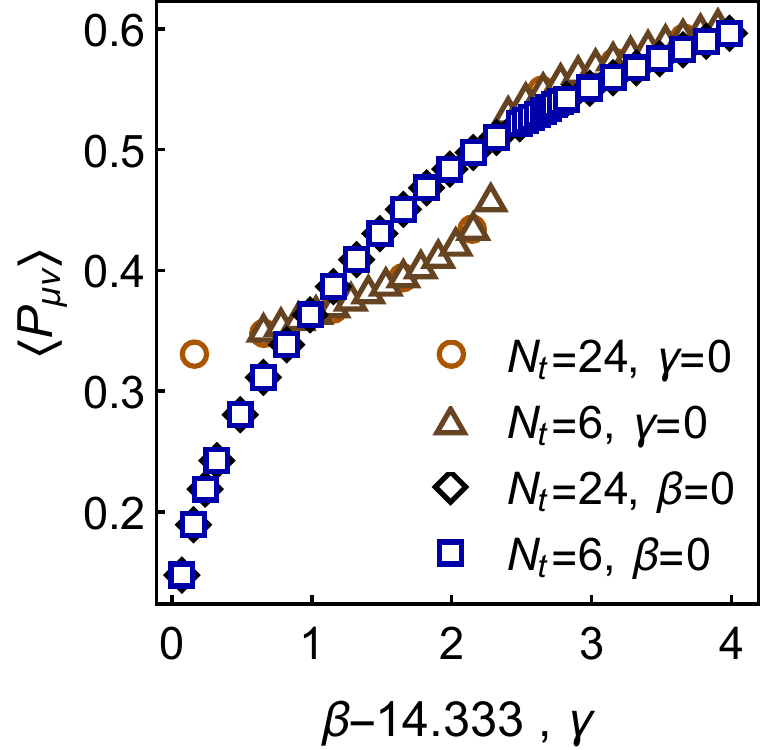}
\end{minipage}\hfill
\begin{minipage}[t]{0.49\linewidth}
\vspace{0pt}
\centering
\includegraphics[height=1.0\linewidth,keepaspectratio,right]{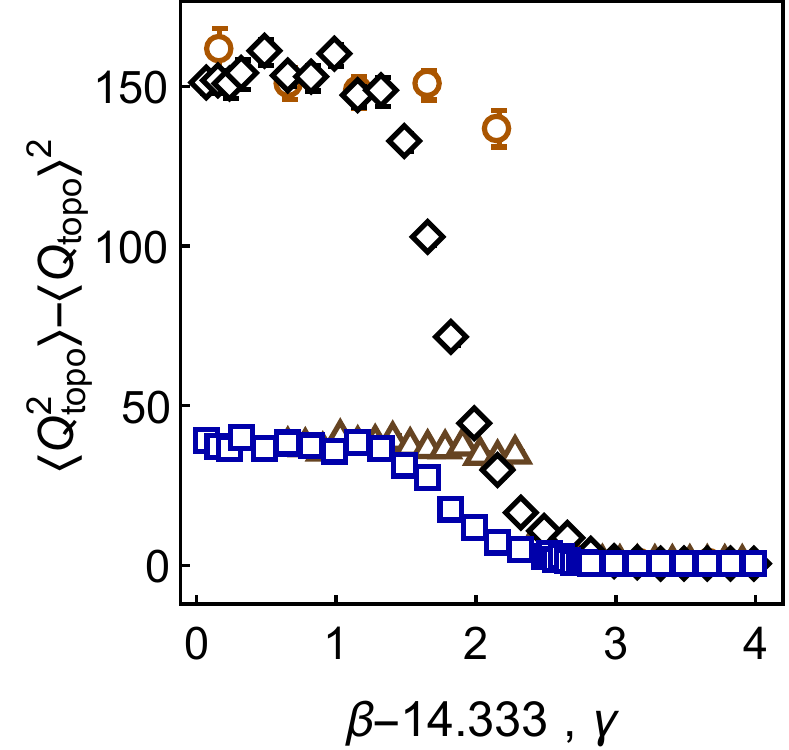}
\end{minipage}
\end{minipage}\hfill
\begin{minipage}[t]{0.51\linewidth}
\centering
\begin{minipage}[t]{0.49\linewidth}
\vspace{0pt}
\centering
\includegraphics[height=1.0\linewidth,keepaspectratio,right]{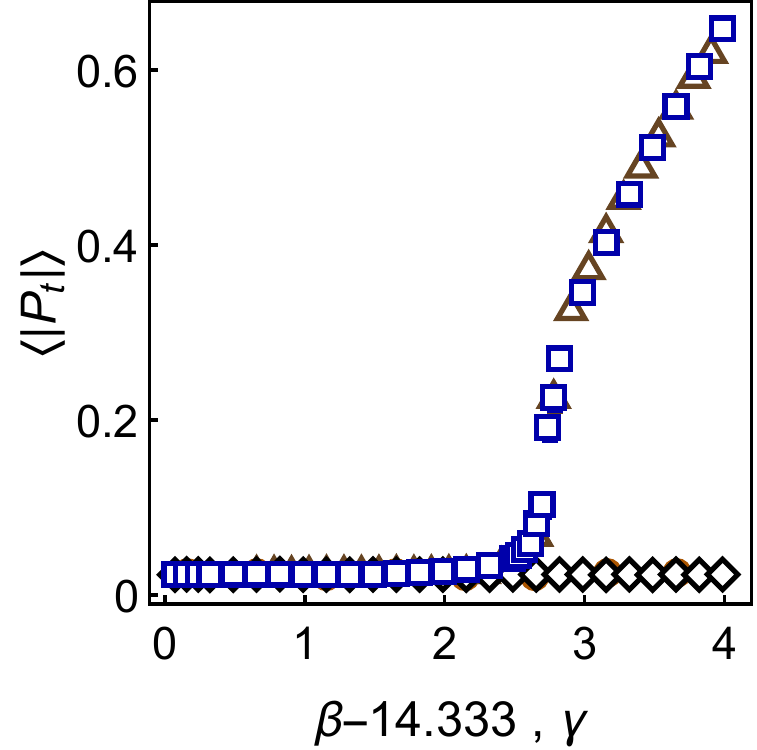}
\end{minipage}\hfill
\begin{minipage}[t]{0.49\linewidth}
\vspace{0pt}
\centering
\includegraphics[height=1.0\linewidth,keepaspectratio,right]{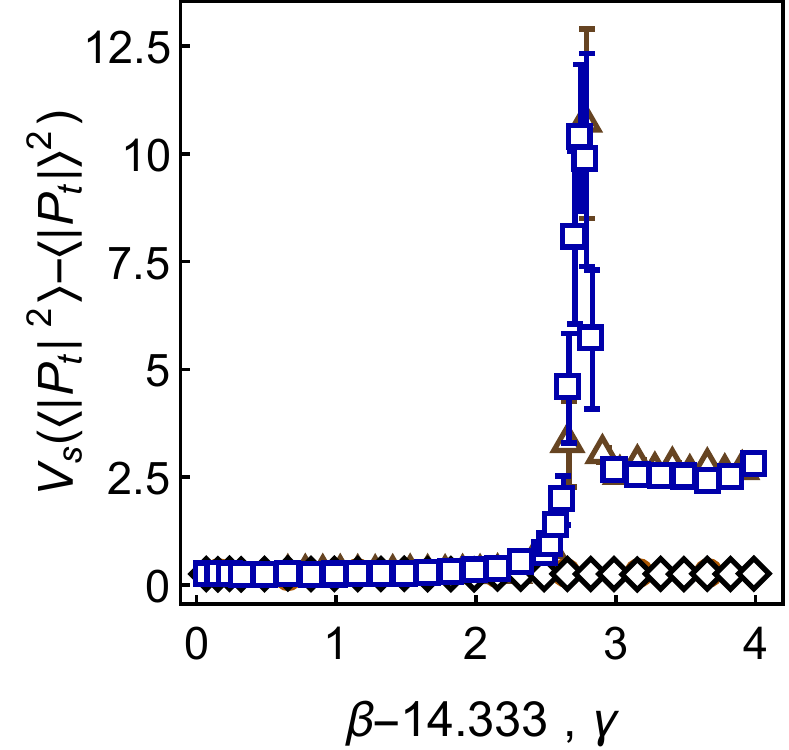}
\end{minipage}
\end{minipage}
\caption{Same as Fig.~\ref{fig:actioncompsu2}, but for pure gauge $\SU{5}$.}
\label{fig:actioncompsu5}
\end{figure}

Fig.~\ref{fig:actioncompsu5} provides the same information as Fig.~\ref{fig:actioncompsu2} but for $\SU{5}$ instead of $\SU{2}$. The shift in $\beta$ to match $\gamma$ at week coupling has been determined to be approximately $43/3\approx14.333$. As in the case of pure gauge $\SU{5}$ the bulk transition of the Wilson gauge action is no longer of 2nd but of 1st order, the average plaquette, shown in the top-left panel of Fig.~\ref{fig:actioncompsu5}, has for the WG action a discontinuity at $\gamma=\beta-43/3\approx 2.3$, while for the BP action, the plaquette is completely continuous as function of the inverse coupling $\gamma$. In the continuum phase, i.e. for $\gamma=\beta-43/3>2.3$, the average plaquette values obtained with the two different actions converge only slowly as function of increasing inverse coupling, while topological susceptibility, average Polyakov loop and Polyakov loop variance agree almost immediately. The finite temperature transition for $N_t=6$ occurs in the continuum phase, at around $\gamma=\beta-43/3\approx 2.9$, so that agreement should be expected. For the WG action, the topological susceptibility seems to jump abruptly at the bulk transitions, while the BP action interpolates more smoothly between the continuum and the asymptotic strong coupling value.

\subsection{$\SU{3}$ with $N_{f}=4$ Wilson flavours}\label{eq:fermionicresults}

\begin{figure}[htbp]
\begin{minipage}[t]{0.51\linewidth}
\centering
\begin{minipage}[t]{0.49\linewidth}
\vspace{0pt}
\centering
\includegraphics[height=1.0\linewidth,keepaspectratio,right]{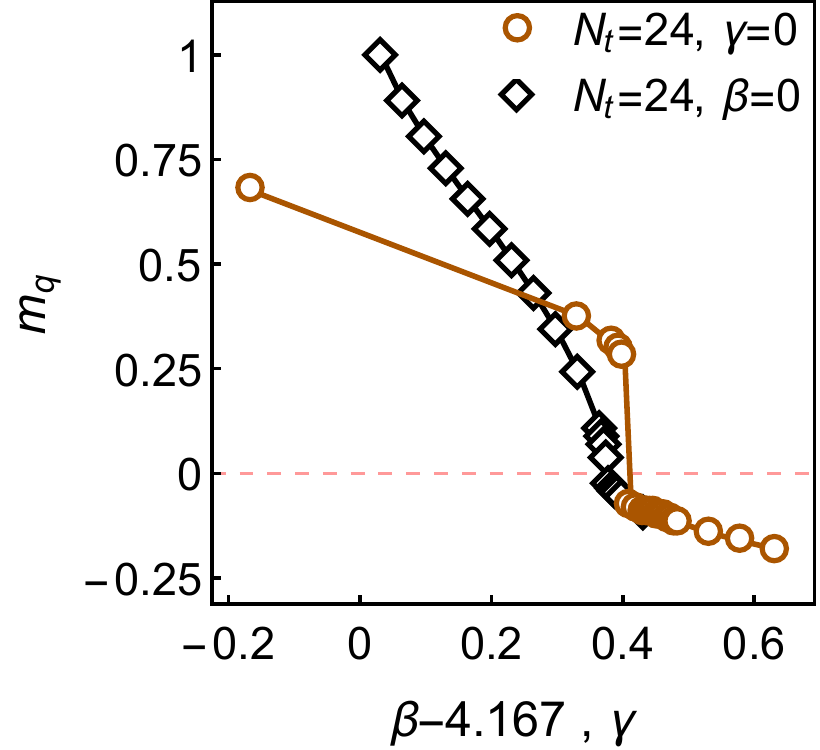}\\[5pt]
\includegraphics[height=1.0\linewidth,keepaspectratio,right]{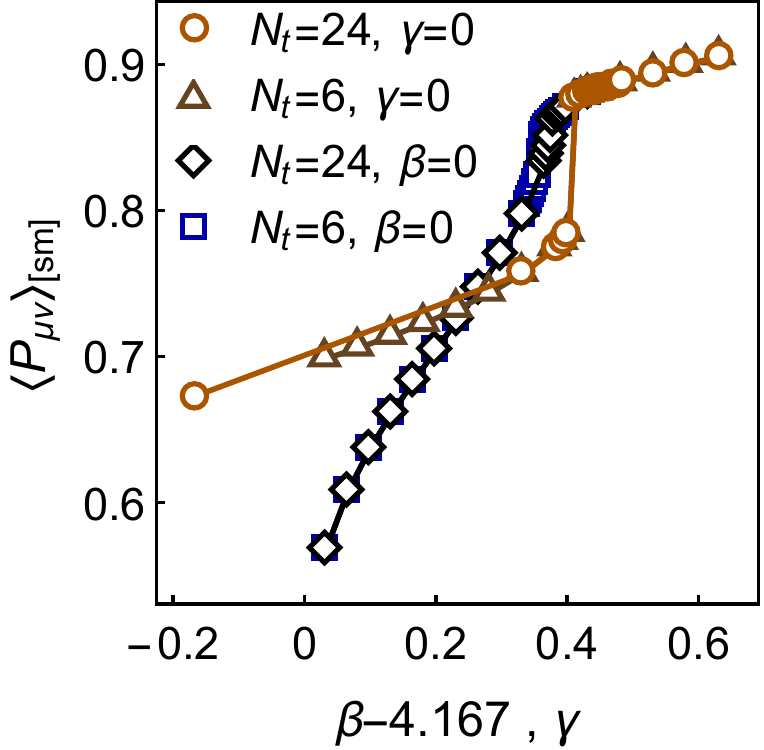}
\end{minipage}\hfill
\begin{minipage}[t]{0.49\linewidth}
\vspace{0pt}
\centering
\includegraphics[height=1.0\linewidth,keepaspectratio,right]{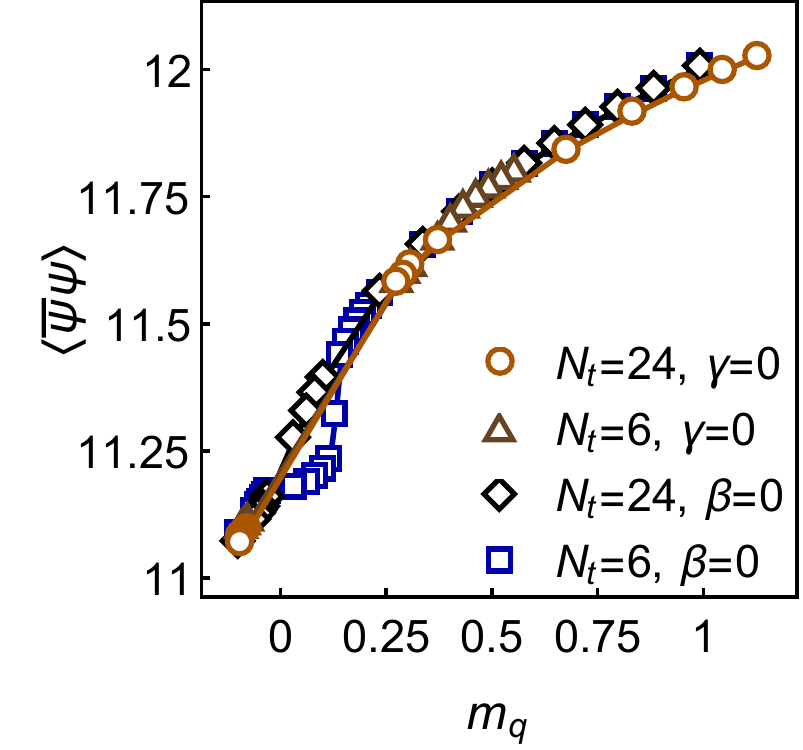}\\[5pt]
\includegraphics[height=1.0\linewidth,keepaspectratio,right]{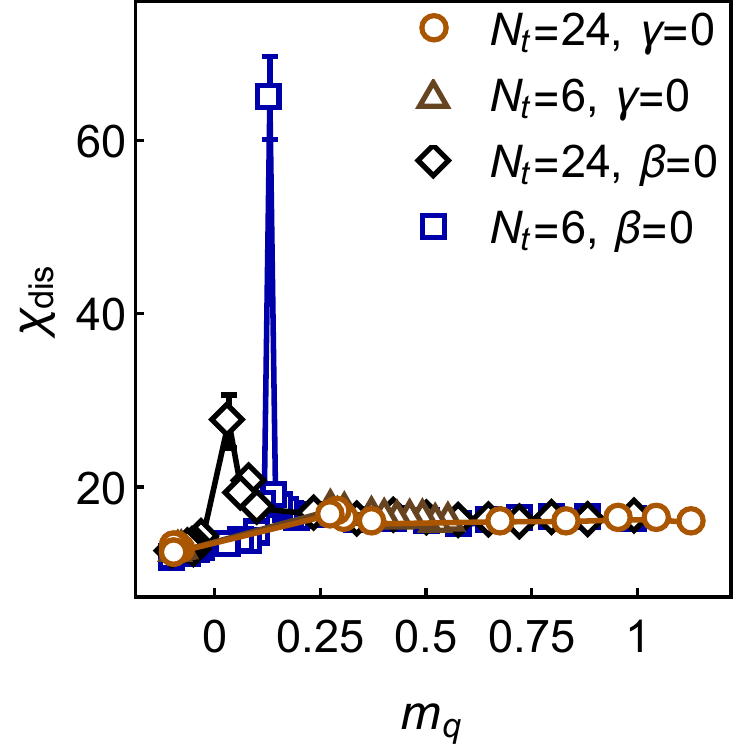}
\end{minipage}
\end{minipage}\hfill
\begin{minipage}[t]{0.51\linewidth}
\centering
\begin{minipage}[t]{0.49\linewidth}
\vspace{0pt}
\centering
\includegraphics[height=1.0\linewidth,keepaspectratio,right]{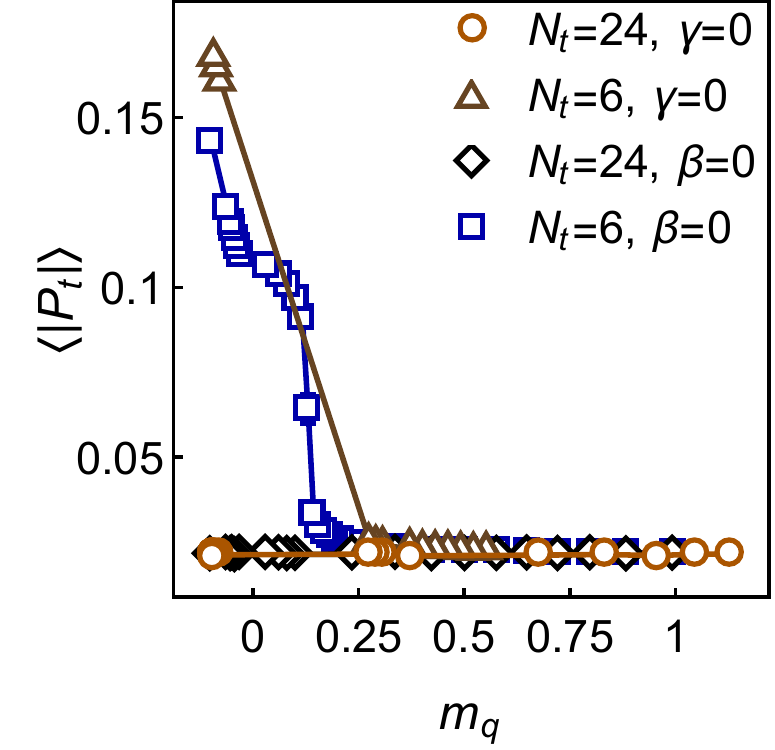}\\[5pt]
\includegraphics[height=1.0\linewidth,keepaspectratio,right]{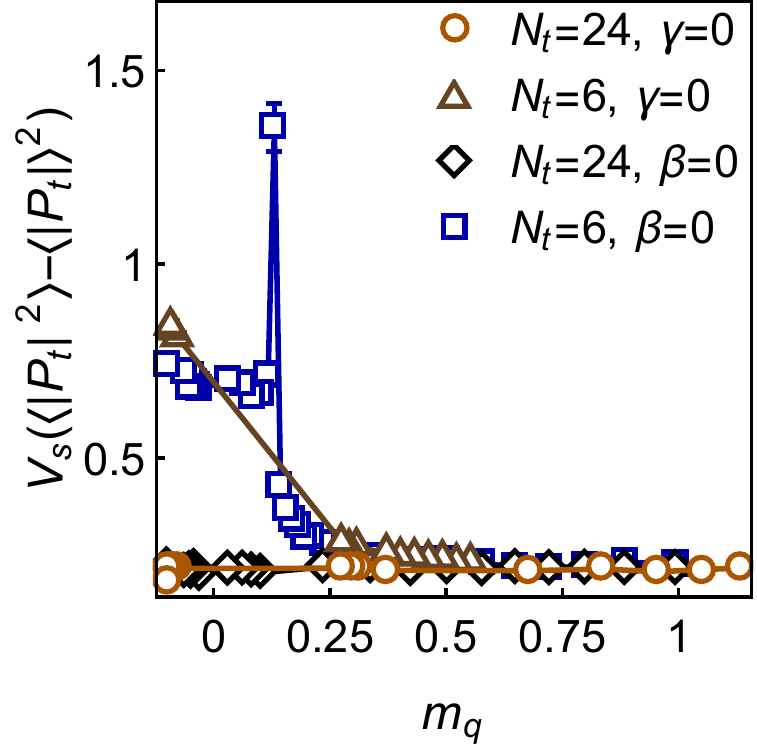}
\end{minipage}\hfill
\begin{minipage}[t]{0.49\linewidth}
\vspace{0pt}
\centering
\includegraphics[height=1.0\linewidth,keepaspectratio,right]{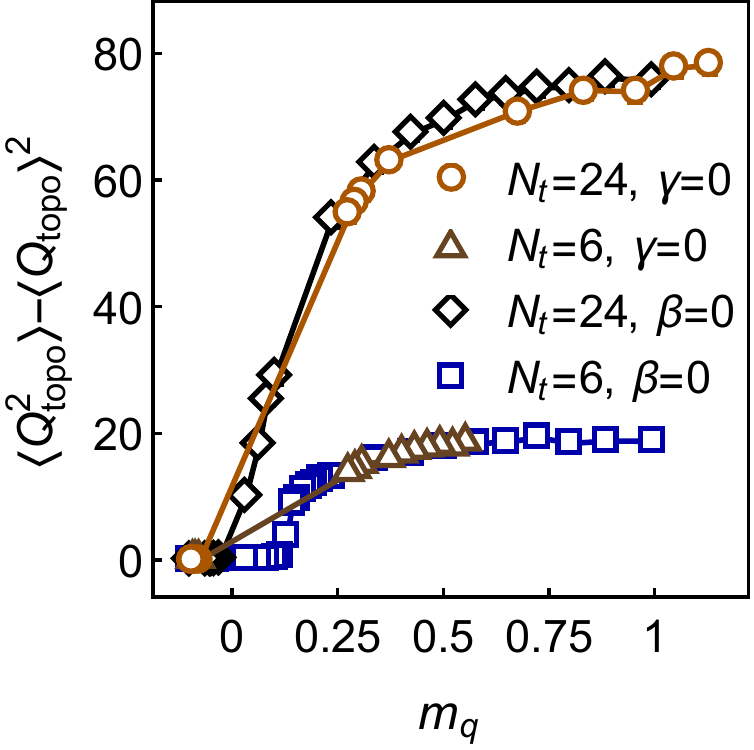}\\[5pt]
\includegraphics[height=1.0\linewidth,keepaspectratio,right]{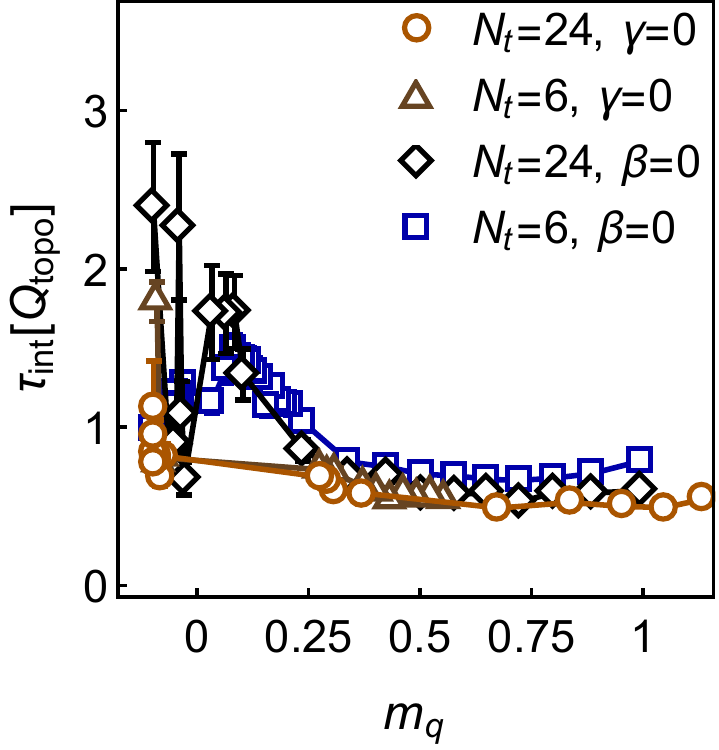}
\end{minipage}
\end{minipage}
\caption{The figure shows simulation results for an $\SU{3}$ lattice gauge theory, coupled via 2-step stout smeared links to $N_f=4$ degenerate Wilson clover fermion flavors with hopping parameter $\kappa=1.358$. As in the previous figures, the orange circles and brown triangles correspond, respectively, to zero and finite temperature results obtained with the Wilson gauge action \eqref{eq:wilsongaugeaction} and the black diamonds and blue squares to corresponding results obtained with the bulk-preventing action \eqref{eq:bpaction}. The first column shows the PCAC quark mass (top) and average smeared plaquette (bottom) as functions of $\gamma$ resp. $\beta-25/6$. The remaining columns show the quantities as functions of the PCAC quark mass (obtained from the $N_t=24$ data). The second column shows the chiral condensate (top) and disconnected chiral susceptibility (bottom), the third column show the temporal Polyakov loop (top) and corresponding variance (bottom), and the last column shows the topological susceptibility (top) and integrated autocorrelation time of the topological charge (bottom).}
\label{fig:actioncompsu3}
\end{figure}

For the $\SU{3}$ gauge theory, the transition between the continuum and bulk phases is usually a cross-over with the Wilson gauge action. However, if the theory is coupled to fermions, it can happen that the transition turns into first order. In Fig.~\ref{fig:actioncompsu3} we show an example of such a case, where the $\SU{3}$ lattice gauge theory is coupled to $N_f=4$ mass-degenerate Wilson-clover fermion flavours via 2-step stout smeared links. The hopping parameter is set to $\kappa=1.358$. As can be seen in the first column of Fig.~\ref{fig:actioncompsu3}, for this value of $\kappa$, there is no value of $\beta$ for which the Wilson gauge action \eqref{eq:wilsongaugeaction} would allow one to sample a physical phase: for $\gamma=\beta-25/6<0.4$ the system is in the bulk phase, and for $\gamma=\beta-25/6>0.4$ the PCAC quark mass, $m_q$, is negative. These two unphysical phases are separated by a first order bulk transition at which the PCAC quark mass jumps from a finite positive to a finite negative value. At finite temperature this bulk transition triggers also the deconfinement transition.

With the bulk-preventing action \eqref{eq:bpaction}, the bulk transition is absent and the PCAC quark mass approaches $m_q=0$ continuously. The small gap in the data around $m_q=0$ is due to the slowing down caused by zero-eigenmodes of the Wilson-Dirac operator when $m_q\to 0$. This could be avoided by using e.g. Schr\"odinger functional boundary conditions.

There are two main observations we would like to point out in connection with Fig.~\ref{fig:actioncompsu3}:
\begin{enumerate}[label=(\alph*)]
\item the presence of the bulk transition prevents the Wilson gauge action from resolving the zero-temperature chiral and finite-temperature deconfinement transition. This is visible in the second and third columns of Fig.~\ref{fig:actioncompsu3}, where the chiral condensate and disconnected chiral susceptibility, as well as temporal Polyakov loop and corresponding variance are plotted as functions of the PCAC quark mass $m_q$. Results from the two gauge actions agree deep in the strong-coupling and deep in the negative mass phase, but due to the bulk-transition, the Wilson gauge action skips the critical $m_q$-values. With the bulk-preventing action on the other hand, there is no discontinuity in $m_q$ and the zero-temperature chiral transition near $m_q=0$, as well as the finite-temperature deconfinement transition at $m_q\approx 0.14$ are resolved.
\item As shown in the last column of Fig.~\ref{fig:actioncompsu3}, also when coupled to fermions,  fluctuations of the gauge-topology are not hindered by using the bulk-preventing action and HMC updating. 
\end{enumerate}

\section{Conclusions \& outlook}

We have identified in Sec.~\ref{sec:avoidinglatbulkphase} a mechanism which is responsible for the formation of bulk-configurations in simulations of lattice $\SU{N}$ gage theories using Wilson's plaquette gauge action. We then proposed a one-parameter family of alternative gauge actions, which possess the same naive continuum limit as the Wilson plaquette gauge action, but which, when used in combination with an HMC update algorithm, prevent the creation of bulk configurations. 

In Sec.~\ref{sec:rsults}, we then tested our bulk-preventing simulation framework for pure gauge $\SU{2}$, pure gauge $\SU{5}$, and for $\SU{3}$ with $N_f=4$ mass-degenerate Wilson-clover fermion flavours with hopping parameter $\kappa=1.358$, and which coupled to the gauge field via 2-step stout smeared link variables. We found that in all three cases, the bulk-preventing action \eqref{eq:bpaction} with $n=2$ does indeed removed the bulk-transition and reproduces at sufficiently weak coupling the same results as the Wilson plaquette action. 
In the case of the fermionic $\SU{3}$ theory, the Wilson gauge action could for the given simulation parameters resolve neither the zero-temperature chiral transitions, nor the finite-tempearture deconfinement transitions, as the bulk transition caused the PCAC quark mass $m_q$ to jump from a large positive to a large negative value, and thereby skipping the whole for the transitions relevant interval. With the bulk-preventing action \eqref{eq:bpaction} on the other hand, the bulk transition was again absent and $m_q$ could approach the chiral limit seemingly continuously. Both, the zero-temperature chiral and the finite-temperature deconfinement transition were nicely resolved.
Noteworthy is also, that the bulk-preventing actions do  not seem to hinder any processes relevant to topology fluctuations.

We have applied the bulk-preventing action in a study of the critical endpoint of the 1st-order deconfinement transition-line in the same $\SU{3}$ + $N_f=4$ Wilson-clover flavour set-up as above. The corresponding publication is currently in preparation.

Further directions to follow include the combination of the Wilson gauge action and different bulk-preventing actions \eqref{eq:bpaction} (i.e. different $n$) and to investigate their impact on finite size effects, as well as inspecting the properties of a gradient flow which is governed by actions of the form \eqref{eq:bpaction}.

\end{document}